\documentclass[aps,prb,twocolumn,superscriptaddress,showpacs]{revtex4}
\usepackage{graphicx }
\usepackage{amsmath}

\begin{document}

\title{Evidence for magnetoplasmon character of the cyclotron
resonance response of a two-dimensional electron gas}

\author{C. Faugeras}
 \affiliation{Grenoble High Magnetic Field
Laboratory, CNRS, B.P. 166, 38042 Grenoble Cedex 9, France}
\author{G. Martinez}
\affiliation{Grenoble High Magnetic Field Laboratory,  CNRS, B.P.
166, 38042 Grenoble Cedex 9, France}
\author{A. Riedel}
\affiliation{Paul Drude Institute, Hausvogteiplatz 5-7, D-10117 Berlin, Germany}
\author{ R. Hey }
\affiliation{Paul Drude Institute, Hausvogteiplatz 5-7, D-10117 Berlin, Germany}
\author{ K. J. Friedland}
\affiliation{Paul Drude Institute, Hausvogteiplatz 5-7, D-10117 Berlin, Germany}
\author{Yu. Bychkov }
\affiliation{Grenoble High Magnetic Field Laboratory,  CNRS, B.P.
166, 38042 Grenoble Cedex 9, France} \affiliation{L. D. Landau
Institute for Theoretical Physics, Academy of Sciences of Russia,
 117940 Moscow V-334, Russia}

\date{\today}

\begin{abstract}
Experimental results on the absolute magneto-transmission of a
series of high density, high mobility GaAs quantum wells are
compared with the predictions of a recent magnetoplasmon theory
for values of the filling factor above 2. We show that the
magnetoplasmon picture can explain the non-linear features
observed in the magnetic field evolution of the cyclotron
resonance energies and of the absorption oscillator strength. This
provides experimental evidence that inter Landau level excitations
probed by infrared spectroscopy need to be considered as many body
excitations in terms of magnetoplasmons: this is especially true
when interpreting the oscillator strengths of the cyclotron
transitions.
\end{abstract}

\pacs{78.67.De, 73.21.Fg, 78.66.Fd}

\maketitle

\section{Introduction}

Many theoretical reports \cite{By,KH,KM,By1} have clearly stated
that excitations between Landau levels (LL) of a two-dimensional
electron gas (2DEG) subjected to a magnetic field B have to be
analyzed, because of many body interactions, as excitonic
transitions which are often referred to as magnetoplasmons (MP).
These excitations display a specific dispersion as a function of
the two dimensional wave vector $\vec{K}$ of the exciton. Because
of Kohn's theorem \cite{Kohn} which states that no manifestations
of electron-electron interactions can be observed in cyclotron
resonance (CR) experiments unless the translational symmetry is
broken or non-local potentials are present (giving rise to
nonparabolicity (NP) effects for instance), it was considered that
the departure from the one-electron model due to the MP response
should hardly be seen experimentaly. However, when NP effects are
present, the theory \cite{KM,By1} predicts a mixing of the MP
modes for small values of $\vec{K}$ which could be, a priori,
visible in experiments. Some signatures of coupled cyclotron
resonance modes have already been observed in the past years
\cite{Ensslin,Hu95} but these works focused on low density samples
and as a consequence on values of the filling factor below 2, a
limit which is nowadays well understood in terms of coupled
transitions in a two-component system, with a coupling even
observed at high temperatures \cite{Hu95}. More recently, Manger
\textit{et al.} \cite{Manger01} studied the cyclotron resonance
response of high density samples similar to the ones presented in
this work, and for integer values of the filling factor. The
observed behavior in this case is different from the previous
studies as the coupling disappears for temperature higher than 1.4
K. These authors show that the one electron picture fails to
interpret the oscillator strength of the split cyclotron resonance
line. In a more recent report \cite{By2}, the conductivity of a
2DEG in the Faraday configuration (the light vector $\vec{q}$
being parallel to the magnetic field $\vec{B}$ and perpendicular
to the plane of the 2DEG) has been calculated in the frame of the
MP picture. We will present in this paper a detailed comparison of
experimental results obtained in a quantitative way with the
predictions of the MP model derived in the Hartree-Fock
approximation ignoring impurity and spin-orbit interactions for
integer and non-integer values of the filling factor. We will show
that the magnetoplasmon picture can explain some of the features
observed in the cyclotron resonance response of a 2DEG.

A CR transition line is characterized by its energy position, its
linewidth and its intensity. Electron-electron interactions being
non-dissipative to first order, the MP should not contribute to
the linewidth of the transition but both the energy position and
the intensity of the CR transitions are expected to be modified.
In the one-electron picture of CR, for non integer values of the
filling factor greater than 2, one expects three distinct
transitions between successive Landau levels which conserve the
spin. One of the predictions of the MP model \cite{By2} is that
the excitonic dispersion is composed of three branches with only
two of them of lower energies being essentially infrared active.
Therefore, whatever are the importance of NP effects in orbital
and spin levels, one should observe, in the Faraday configuration,
\textit{at most only} two lines resulting from the mixing of the
three one-electron transitions. Because of this mixing, the
oscillator strengths of the transitions are significantly modified
with respect to those deduced from the one-electron model. The
third MP transition which is practically infrared inactive is
shifted to higher energies by an amount which depends on the
exchange interactions between different LL. Though not directly
measurable, these interactions manifest themselves in the magnetic
field variation of the CR energies which develops ``kinks`` at odd
integer values of the filling factor $\nu = n_{S} \phi_0/B$
($\phi_0$ being the flux quantum and $n_{S}$ the areal density of
carriers) for $\nu>2$.

\section{Experimental details}

New refined experiments have been performed on single doped
 GaAs quantum wells (QW) already used in a previous report
 \cite{Faugeras}. The GaAs QW is sandwiched between two GaAs/AlAs
 superlattices with the Si-n-type doping performed symmetrically on
both sides of the QW \cite{Friedland}. The whole epistructure has
been lifted-off from the growth substrate and deposited on an
insulating Si wedged substrate which is transparent in the Far
Infrared range of energy to avoid the strong optical phonon
absorption (reststrahlen band) which occurs in traditional GaAs
thick substrates. We have performed magneto-infrared transmission
measurements in the Faraday configuration with unpolarized
infrared light using a Bruker Fourier Transform Spectrometer
IFS-113 which was connected to a resistive 28 T magnet via a
metallic waveguide. All measurements were made at a fixed
temperature of 1.8 K. We use a rotating sample holder which allows
us to switch in-situ between the sample and a reference sample.
For each value of the magnetic field, we have measured the
transmission of the sample and of a reference (a silicon
substrate) to normalize the transmission. These measurements are
thus absolute measurements and exact absorption amplitudes can be
extracted for the comparison with models. Moreover, as optical
phonons in the AlAs/GaAs short period superlatice and in the GaAs
QW also present infrared signatures we were led to normalize the
absolute transmission at a given magnetic field B by the absolute
transmission measured at B=0 T. In the following, these absolute
spectra normalized to the B=0 T spectra will be refered to as
relative transmission spectra and will then be compared to a
simulation \cite{By3} of the multilayer transmission taking into
account all the different dielectric layers of the sample. The
whole procedure minimizes the errors due to experimental or/and
dielectric artefacts.

In this paper, we will present results obtained for three samples
labelled 1038,1201 and 1211, with a typical size of 4 x 4 mm. The
sample 1038 corresponds to a QW width L=10 nm, a carrier density
$n_{S}= 12.8\times 10^{11}cm^{-2}$ and a mobility of $\mu_{DC}=
114$ $ m^{2}V^{-1}s^{-1}$ whereas for the samples 1201 and 1211,
originating from the same wafer, L =13 nm, $n_{S}= 9.4\times
10^{11}cm^{-2}$ and $\mu_{DC}= 280$ $m^{2}V^{-1}s^{-1}$. These
concentration and mobility values are obtained from transport
measurements performed on parent samples without the lift-off
process.

\section{Results}

In order to fit the measured infrared transmission, one has to use
the components of the dielectric tensor $\overline{\varepsilon}$
of the doped QW. As shown in Ref. \cite{By2}, these components can
be formally written as:
\begin{align}
\varepsilon_{\parallel}&=\varepsilon_{L}-\frac{(\omega_{p})^{2}}{\omega}
\sum_{p}\frac{I_{p}\hspace{0.1cm}(\omega+\imath\delta_{p})}
{(\omega+\imath\delta_{p})^{2}-(E_{p}^{MP})^{2}}\notag\\
\varepsilon_{\perp} &= -\imath\frac{(\omega_{p})^{2}}{\omega}
\sum_{p}\frac{I_{p}\hspace{0.1cm}
E_{p}^{MP}}{(\omega+\imath\delta_{p})^{2}-(E_{p}^{MP})^{2}}
\end{align}

where $\varepsilon_{L}$ is the lattice contribution to the
diagonal part of the tensor, $(\omega_{p})^{2}= 4\pi
n_{S}e^{2}/(Lm^{*})$ is the square of the plasma frequency defined
for a mean value of the carrier effective mass $m^{*}$. One can
note that formally Eq.(1) is very similar to the standard
expression derived from the Drude model. However here the
significance of the parameters is different: the summation over p
(= 0,1) corresponds to the two CR active MP modes with an energy
$E_{p}^{MP}$, and a corresponding oscillator strength $I_{p}$.
Both oscillator strength depend on $K_{\parallel}=
\mid\vec{K}\mid$ but are used in the fitting procedure as
additional parameters with the constraint that $\sum_{p}I_{p}=1$.
The $K_{\parallel}$-dependence of $I_{p}$ is relatively weak but
its value at $K_{\parallel}\approx 0$ is significantly different
from the value obtained in the one-electron model due to the
mixing of the one-electron transitions implied in the MP approach
\cite{By2}. In addition the damping parameter $\delta_{p}$ is used
to fit the width of each transition.

To illustrate this point, we compare in Fig.1 the experimental
spectrum for sample 1211 measured at $\nu=2.34$, with the
simulated spectra using Eq.1 for two distinct approaches: in
Fig.1a  the simulation, assuming that the oscillator strength is
given by the one-electron model, uses four independent fitting
parameters (the energies $E_{0}^{MP}$, $E_{1}^{MP}$, and the
respective damping parameters $\delta_{0}$, $\delta_{1}$); in
Fig.1b an additional fitting parameter ($I_{0}$) is used to
simulate the spectrum. It is clear from the results, that using
the one-electron model to deduce the carrier concentration of a
2DEG is not correct and in the present case, leads to a
concentration which is $6-8 \%$ lower than the one deduced from
transport measurements. This result is of particular importance
when interpreting CR data.

\begin{figure}
\includegraphics*[width=0.9\columnwidth]{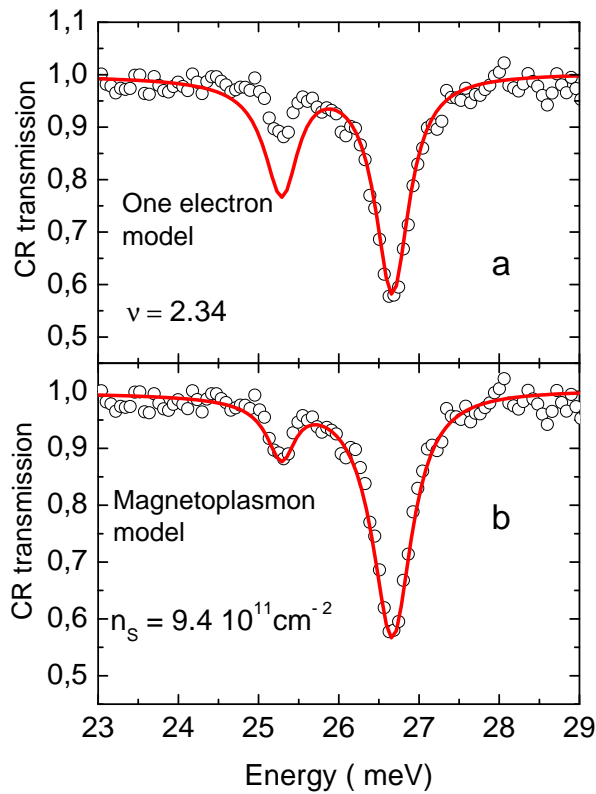}
\caption{\label{Fig.1} Comparison of CR experimental transmission
spectra (open circles) and theoretical simulated spectra (full
lines) of sample 1211 for $\nu=2.34$. In panel a) the fit is made
using the oscillator strength given by the one-electron model
whereas in panel b) the oscillator strength of the transitions is
fitted (see text).}
\end{figure}

Fig.2 and Fig.3 display the results of the fitting procedure
obtained for samples 1201/1211 and 1038 respectively. In the upper
panels the data relative to the damping parameters $\delta_{0}$
and $\delta_{1}$ are reported. On physical grounds, besides a weak
constant value of these parameters reflecting the small residual
imperfections of the sample, any deviation from this value
reflects the presence of an additional dissipative interaction. In
order to have a more detailed vision of the variations of energies
with the filling factor, the quantities displayed in the lower
panel of Fig.2 and Fig.3 are $E_{0}^{MP}/B$ (open circles) and
$E_{1}^{MP}/B$ (crosses). In the Faraday configuration, the wave
vector of the light $\overrightarrow{q}$ is parallel to
$\overrightarrow{B}$, and the only non-zero theoretical component
of $\overrightarrow{q}$ is $q_{z}$. Therefore, in principle,
$K_{\parallel}$ should be $0$. In practice however one has to face
the divergence of the beam which can be important especially when
using light-pipes to transfer the electro-magnetic field. When the
full multi-dielectric treatment is performed \cite{By3} to
evaluate the transmission of such samples, values for $q_{z}= 4\pi
Re(\widetilde{n})/\lambda$ (where $\widetilde{n}$ is the complex
value of the index of refraction and $\lambda$ is the wave-length
corresponding to the CR transition) are found to range between 4
and $7\times 10^{4}cm^{-1}$. As a consequence, allowing for
divergence of the infrared beam leads to consider values of
$q_{\parallel}=K_{\parallel}\approx 10^{4}cm^{-1}$. We display in
the lower panels of Fig.2 and Fig.3 the evolutions for both
transitions expected from the MP model \cite{By2} for
$K_{\parallel}= 1\times10^{4}cm^{-1}$ (full and dashed lines). Of
course these values vary with the magnetic field and the curves
displayed in these figures have to be understood as a frame
surrounding the corresponding variations.

As already reported \cite{By2,By3}, one has to use an independent
model to reproduce NP effects in the MP approach and the one used
here is that given in Ref.\cite{Hermann} with a fitting parameter,
the QW effective gap, adjusted to fit the data for $E_{0}^{MP}/B$
at $\nu= 3$. Changing this parameter does not modify the shape of
the magnetic field evolution of the $E_{p}^{MP}/B$ and only
results in a rigid shift \cite{comment}. One notes that, for both
samples, the variations of $E_{0}^{MP}/B$ as a function of $\nu$
display a kink at $\nu =3$ which is well reproduced by the MP
model. The discontinuity of $E_{0}^{MP}/B$ observed for the sample
1038 (Fig.3) at $\nu \approx 2.4$ occurs, within the experimental
error, for $E_{0}^{MP}$ equals the energy $\hbar\omega_{TO}$ of
the TO phonon mode of GaAs. This discontinuity and the related
increase of $\delta_{0}$, also observed in samples 1201/1211 but
at $\nu< 2$, reflects an interaction previously reported for a
doped GaAs QW \cite{Faugeras} as well as for a  doped GaInAs QW
\cite{Faugeras1}. This interaction which apparently concerns
$E_{0}^{MP}$ and not $E_{1}^{MP}$ is, at present, not well
understood and will not be discussed here. One notes that the
variation of $\delta_{0}$ is quite smooth as long as $E_{0}^{MP}$
is well below $\hbar\omega_{TO}$. On the other hand, the variation
of $\delta_{1}$ goes through a pronounced maximum at $\nu \approx$
2.5 and 4.5. This reflects, as already pointed out, the presence
of an additional dissipative interaction mechanism which mainly
concerns $E_{1}^{MP}$.

\begin{figure}
\includegraphics*[width=0.9\columnwidth]{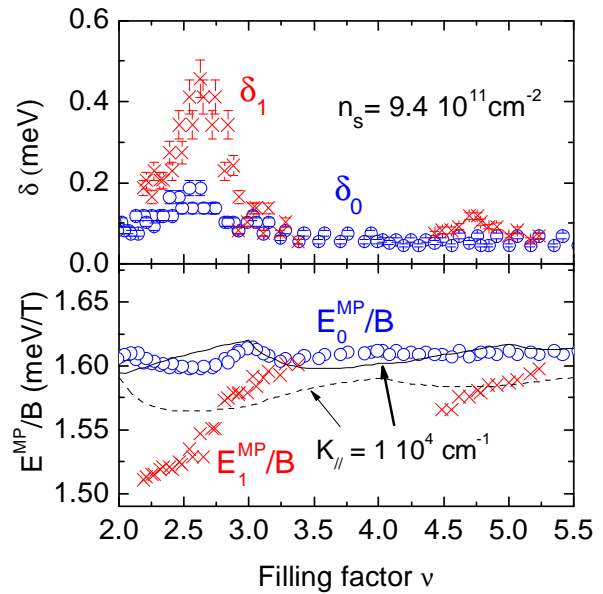}
\caption{\label{Fig.2} Collected data for samples 1201/1211. Top
panel: variation as a function of $\nu$  of the fitted damping
parameters $\delta_{0}$ (open circles) and $\delta_{1}$ (crosses)
of the CR transitions. Bottom panel: comparison between the
experimental results for $E_{0}^{MP}/B$ (open circles) and
$E_{1}^{MP}/B$ (crosses) with the predictions of the MP model for
$K_{\parallel}= 1\times 10^{4}cm^{-1}$: full line corresponds to
$E_{0}^{MP}/B$ and dashed line to $E_{1}^{MP}/B$.}
\end{figure}

\begin{figure}
\includegraphics*[width=0.9\columnwidth]{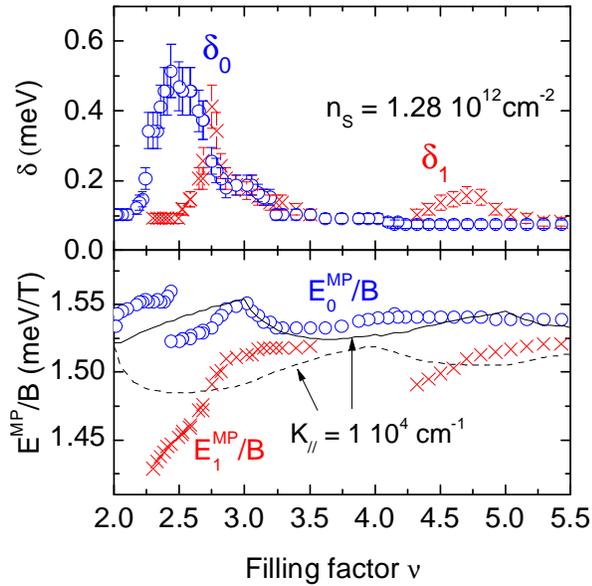}
\caption{\label{Fig.3} Sample 1038. Top panel: Variation as a
function of $\nu$ of the experimental damping parameter
$\delta_{0}$ (open circles) and $\delta_{1}$ (crosses) of the CR
transitions. Bottom panel: comparison between the experimental
results for $E_{0}^{MP}/B$ (open circles) and $E_{1}^{MP}/B$
(crosses) with the predictions of the MP model for $K_{\parallel}=
1\times 10^{4}cm^{-1}$: full line corresponds to $E_{0}^{MP}/B$
and dashed line to $E_{1}^{MP}/B$.}
\end{figure}

To complete the report on the fitting procedure, the oscillator
strength $I_{0}$ of the $E_{0}^{MP}$ transition is displayed in
Fig.4 as a function of $\nu$ for all samples. The corresponding
oscillator strength for $E_{1}^{MP}$ is the complement to 1 of
this variation. The fitted oscillator strength goes through a
minimum at $\nu =$  3 and 5: the relative depth of these minima
depends on $\nu $ for each sample and also on the carrier
concentration.  Here the data are compared to the predictions of
the MP model for different values of $K_{\parallel}$. For any
small non zero value of $K_{\parallel}$, $I_{0}$ rises to 1 for
$\nu$ values close to an even integer.

In the following we will discuss the experimental results obtained
in the framework of the MP picture and they will be compared to
the expectations of the one-electron picture. First, the evolution
of $E_{0}^{MP}/B$ is quite constant for $ 2< \nu <6$ which is not
predicted by the one electron model including NP effects  but well
reproduced by the MP model. Second, the kink observed clearly for
$\nu = 3$ is also well reproduced. This is a fundamental point
because, as shown in Ref.\cite{By2}, the presence  of this
``kink`` is due to the change of the exchange energies across odd
filling factors larger than 2 and is a direct signature of the MP
effects. A good agreement is also observed for $I_{0}$ as far as
the minima at odd integer filling factors are concerned as well as
their relative variations on $\nu $ and $n_{S}$. This is in
contrast to the predictions of the one-electron model in which not
only minima of $I_{0}$ do not occur at odd integer values of $\nu
$ but are also independent on $\nu $ and $n_{S}$. One notes in
addition that the results discussed in Fig.1 become coherent in
the MP picture, because the use of carrier densities deduced from
the one-electron model would lead to the kink of $E_{0}^{MP}/B$
occurring at filling factors around 2.8 where no physical
arguments could justify such a discontinuity. Therefore it is
clear from these data that the MP picture should be considered
when interpreting CR experiments. It is important to note here
that, whereas the "kink" feature appearing in the energy vari-
ation of the $E_{0}^{MP}/(B)$ has an amplitude decreasing when
$n_S$ decreases, the characteristic extrema of the oscillator
strengths at odd filling factors is a robust signature of the MP
character of the CR transitions which, in practice, does not
depend on the carrier concentration.

There are however apparent discrepancies between the reported data
and the predictions of the MP model. First the evolution of
$E_{1}^{MP}$ (Fig.2 and Fig.3) is not reproduced by the model,
especially for $\nu$ values between 2 and 3, and to a less extent
between 4 and 5. In the same range of filling factors, the
variation of the oscillator strength (Fig.4) is also not correctly
taken into account (because of the sum rule $\sum_{p}I_{p}=1$, the
discrepancy for $I_{0}$ is the consequence of an overestimation of
$I_{1}$ by the MP model). Further refinements of the MP model,
such as changing the NP model \cite{Ekenberg,Golubev} or
introducing spin-orbit interaction \cite{By4}, were unable to
explain, even qualitatively, the observed discrepancies.

\begin{figure}
\includegraphics*[width=0.9\columnwidth]{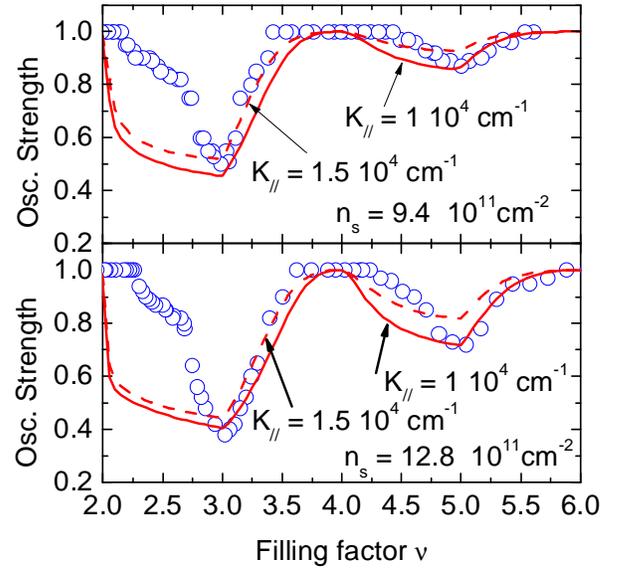}
\caption{\label{Fig.4}  Variation as a function of $\nu$ of the
fitted oscillator  strength $I_{0}$ of the $E_{0}^{MP}$ transition
(open circles) for samples 1201/1211 (top panel) and sample 1038
(bottom panel). These variations are compared to the predictions
of the MP model for $K_{\parallel}= 1\times 10^{4}cm^{-1}$ (full
lines) and $=1.5 \times10^{4}cm^{-1}$ (dashed lines).}
\end{figure}

One can note however that these discrepancies occur when the
dissipative mechanism is switched on. Whatever is its origin, the
occurrence of such a mechanism will lead to a self-energy
$\Sigma(\omega)= Re(\Sigma(\omega))+\imath Im(\Sigma(\omega))$
which will modify the poles of $\overline{\varepsilon}$. In the
frame of the response theory $Im(\Sigma(\omega))$ and
$Re(\Sigma(\omega))$ are related each other by the
Kramers-Kr\"{o}nig (KK) relations. Assuming that
$Im(\Sigma(\omega))$ is reflected by the observed increase of
$\delta_{1}$ which could be fitted with a Lorenztian curve, the
corresponding KK transform develops an "S" shape variation around
the maximum of the Lorenztian curve and can explain quantitatively
the discrepancies between the experimental and calculated values
of $E_{1}^{MP}/B$ in the range of corresponding filling factors.
This argument has however to be considered as an empirical
approach to this problem: it does explain, \textit{a priori}, the
discrepancies observed for the oscillator strength.

The presence of the dissipative interaction has to be introduced
into the MP model in a self-consistent way. To explain the origin
of this interaction, it is natural to invoke the effects of
impurities which in the present case of high mobility samples
should essentially contribute to long range interactions. The
theoretical approach of the MP model including these effects is
difficult and has only be treated, formally, in the case of
integer filling factors \cite{KH} where, in the present
experimental investigation, nothing spectacular occurs. This
approach reveals however that the introduction of such effects
leads to an additional modification of the mixing of the wave
functions: we therefore believe that a complete treatment of the
MP model, introducing impurity interactions, should  reproduce
self-consistently the observed variation of $E_{1}^{MP}/B$ and
explain the discrepancies observed in the oscillator strengths.

\section{Conclusion}

In conclusion, careful far infra-red magneto-transmission
measurements performed on a series of doped GaAs QW reveal many
specific features in the evolution of the CR energies and of their
oscillator strengths which can be understood in the frame of the
MP picture developed without any additional dissipative
interaction. This proves that cyclotron resonance excitations in
semiconductors are indeed magnetoplasmon excitations due to
many-body interactions, which have to be taken into account when
analyzing cyclotron resonance experiments.

\section*{Acknowledgments}
The GHMFL is "Laboratoire conventionn\'{e} à l'UJF et l'INPG de
Grenoble". The work presented here has been supported in part by
the European Commission through the Grant RITA-CT-2003-505474.


\begin{references}
\bibitem{comment} We adjust this parameter at $\nu=3$ in order to compare
the theoretical and experimental shapes of the "kink".
\end{references}
\end{document}